\documentstyle[11pt]{article}

\newcommand{\be}{\begin{equation}}
\newcommand{\ee}{\end{equation}}
\newcommand{\ba}{\begin{array}}
\newcommand{\ea}{\end{array}}
\newcommand{\bc}{\begin{center}}
\newcommand{\ec}{\end{center}}
\newcommand{\bi}{\begin{itemize}}
\newcommand{\ei}{\end{itemize}}

\newcommand{\disregard}[1]{{}}

\def\bild#1\over#2{\mathrel{\mathop{\kern0pt #1}\limits_{#2}}}

\begin{document}
\centerline {{\bf RANDOM MAGNETIC IMPURITIES AND THE $\delta$ IMPURITY
PROBLEM  \rm}}
\vskip 2cm
\centerline{{\bf Jean DESBOIS, Cyril FURTLEHNER\rm } and
{\bf St\a'ephane OUVRY \rm }\footnote{\it  and
LPTPE, Tour 12, Universit\'e Paris  6 / electronic e-mail:
OUVRY@FRCPN11.IN2P3.FR}}
\vskip 1cm
\centerline{{Division de Physique Th\'eorique \footnote{\it Unit\a'e de
Recherche  des
Universit\a'es Paris 11 et Paris 6 associ\a'ee au CNRS},  IPN,
  Orsay Fr-91406}}
\vskip 3cm
{\bf Abstract : One  considers
the effect of disorder on the 2-dimensional density of states of
an electron in  a constant magnetic field superposed onto a Poissonnian random
distribution of point
vortices.
If one restricts the electron Hilbert space to the lowest Landau
level of the total average
magnetic field, the random magnetic impurity problem
is mapped onto a contact $\delta$ impurity problem.  A brownian motion
analysis of the model, based on brownian probability distributions for
arithmetic area winding sectors, is also proposed. }

\vskip 0.5cm


IPNO/TH 94-77

September 1994

\vfill\eject

{\bf Introduction :}
One considers a 2-dimensional model for an electron of electric
charge $e$ and of mass $m$
subject to  a random magnetic field.
By random magnetic field, one means a constant external magnetic field
$B$ superposed onto a random distribution $\{\vec r_i, i=1,2,....,N\}$
of fixed infinitely thin vortices carrying a flux
$\phi$, modeling magnetic impurities, and characterized  by
the dimensionless Aharonov-Bohm ($A-B$) coupling
$\alpha=e\phi/2\pi=\phi/\phi_o$.
Particular interest
will  be paid  to the effect of disorder on the energy level density
$\rho(E)$
of
the test particle
averaged over the random position of the vortices \cite{0}.
In the thermodynamic limit  $N\to
\infty, V\to\infty$ for a distribution of vortices of
density $\rho=N/V$, the average magnetic field,
$<B>=\alpha \rho\phi_o$, becomes meaningful
in the limit $\rho\to \infty,\alpha\to 0$, with
$\rho\alpha$ kept finite.
However,
as soon as $\rho$ is
finite, and $\alpha$ non vanishing,
corrections due to
disorder should  exhibit non trivial magnetic
impurity signatures, like broadening of Landau levels and localization.

In a first approach to this problem, a brownian motion analysis partially
relying on lattice
numerical
simulations, will exhibit  a global shift $|e<B>/2m|=<\omega_c>$ in the Landau
spectrum of the
average magnetic field. In the case $\alpha=\pm 1/2$, on the other hand,
where the disorder is clearly non perturbative,
a depletion of states at the bottom of the spectrum will
manifest itself by a Lifschitz tail in the average density of states.

Secondly, a quantum mechanical formulation will be used, where short range
singular $A-B$ interactions are properly taken into account by a
wave function redefinition, allowing for an analytical averaging on
the disorder. If one assumes that the total average
magnetic field $B+<B>$ is strong enough so that one can neglect
the coupling between
the lowest Landau level (LLL) and the excited Landau levels by
the random component of the vortex distribution, the
system will  be best described when projected in the LLL.
Since one  has in view a sufficiently
dilute gas of electrons compared to the  available
quantum states in the LLL -the fractional Quantum Hall
regime-, such a restriction is licit. The system will be explicitly
mapped on an
equivalent problem of random $\delta$ impurities, from which additional
information on Landau levels broadening will be extracted.

{\bf Brownian motion :}
One starts from a square lattice of $\cal{N}$ squares of size
$a^2$, in which one randomly drops point magnetic impurities. Let $N_i$ be the
number
of vortices dropped on square $i$. A random configuration $\{N_i\}$ will be
realized with the probability
\be\label{prob} P(\{N_i\})={{N}!\over
{\cal{N}}^{{N}}\prod_{i=1}^{{\cal{N}}} N_i!}\to_{{\cal{N}}\to\infty}
\prod_i^{{\cal{N}}}
{(\rho a^2)^{N_i}e^{-\rho a^2}\over N_i !}\ee
with ${N}/{\cal{N}}=\sum_{i=1}^{{\cal{N}}} N_i/{\cal{N}}=\rho a^2$.
In order to compute the average level density $<\rho(E)>$, one focuses,
in the thermodynamic limit $\cal{N}\to \infty$, on the one-electron
partition function
(in a brownian motion approach  one sets $e=m=1$)
\be\label{PART} Z=Z_o <e^{i\sum_{i=1}^{\cal{N}}  2\pi n_i
N_i\alpha}>_{\{C,N_i\}}\ee
where $\{C\}$ is the set of $L$ steps closed random
walks, and
 $n_i$ is the number of times the square $i$ has been wound around by a
given
random walk in $\{C\}$, i.e. its winding number. $Z_o={1\over 2\pi
t}$ is the free partition
function. (\ref{PART}) is invariant when $\alpha$ is shifted by an
integer, so one can always restrict to $|\alpha|<1$.
Averaging $Z$ with (\ref{prob}) one gets
\be\label{PART'} Z=Z_o<e^{\rho\sum_nS_n(e^{i 2\pi \alpha n}-1)}>_{\{C\}}\ee
where $S_n$ stands for the arithmetic area of the $n$-winding sector of
a given random walk  in $\{C\}$.

Consider first the limit of no disorder, $\alpha\to 0,
\rho\to\infty$, and $<B>$ finite. One expects
\be Z\to Z_o<e^{i<B>\sum_n nS_n}>_{\{C\}}\ee
i.e. the partition function of one electron in an uniform magnetic field
$<B>$.
However, this expectation is not quite correct because of possible
corrections coming from the exponent $\exp(i2\pi n\alpha)-1$. Due to the
non-differentiability of
brownian paths,
a quantity such as $\sum_n
n^2S_n$ is not defined for a typical brownian curve where
$<S_n>={t\over 2\pi n^2}$ \cite{cdo}, with $t$  the length of the curve
($2t=L a^2$). A more precise information on brownian motion winding
probability distributions is necessary. It has been shown \cite{werner}
that $S_n$ is strongly peaked when $n\to
\infty$, namely the probability distribution of the variable
$ x=n^2 S_n$, when
 $n\to\infty$, tends to $\delta(x-{t/2\pi})$,
with a variance smaller than a constant times $n^{-\eta}, \eta=18$. Also,
$S_n$ and $S_{n'}$ are very weakly correlated when $n$ and $n'$ become
large.
Since one wants to compare the partition function $Z$ of the electron
to
its partition function
in the average magnetic field $<B>$, one can as well
consider the partition function $Z_B$ of the electron
not only subject to the random vortices, but
also  to an uniform magnetic field $B$,
\be\label{SA} Z_B=Z_o<e^{-2<B>{\cal{S}}}\cos(<B>{\cal{A}})>_{\{C\}}\ee
where the variables $\cal{S}$ and $\cal{A}$ are defined as
\be \label{sum} {\cal{S}}={1\over 2\pi\alpha}\sum_nS_n\sin^2(\pi\alpha
n)\quad <{\cal{S}}>={\alpha\over|\alpha|}{t\over 4}(1-|\alpha|) \ee
\be {\cal{A}}={1\over 2\pi\alpha}\sum_nS_n(\sin(2\pi\alpha n)+2\pi\alpha
n {B\over <B>})\quad  <{\cal{A}}>=0 \ee
and, in the case $B=-<B>$,  compare $Z_{-<B>}$ to $Z_o$.
For averaging (\ref{SA}) in $\{C\}$, the probability
distributions of $\cal{S}$ and $\cal{A}$ are needed. If one splits the sum in
(\ref{sum}) as $\sum_n=\sum_{|n|\le n'}+\sum_{|n|>n'}$, with $n'$
sufficiently large so that the $\delta(x-t/2\pi)$ probability distribution for
the
variable $x=n^2S_n$
can be used when $|n|>n'$, one gets
\be P({\cal{S}})\to_{\alpha\to 0} \delta({\cal{S}}-{\alpha\over|\alpha|}{t\over
4})\ee
\be P({\cal{A}})\to_{\alpha\to 0}\delta({\cal{A}})\ee
One deduces in the limit $\alpha\to 0$
\be \label{sh}Z_{-<B>}\to_{\alpha\to 0} Z_oe^{-t|<B>|/2}\ee
implying that the system of random vortices is equivalent, as expected, to an
uniform
magnetic field $<B>$, but with an additional positive shift
$|<B>|/2=<\omega_c>$ in the Landau
spectrum.
The origin of this shift can be traced back to the $S_n$ brownian
law. When one counts probability of winding around fixed points via $A-B$
path integral technics \cite{cdo}, the quantum $A-B$ particle is
forbidden to
coincide with the vortex location, via contact repulsive interactions. This
procedure
\cite{acad} has deep implications, both in the $A-B$ and in the
anyon
contexts (many-body $A-B$ problem),  in
particular for Bose-based perturbative  computations.
One will come back to this point in the sequel. For the time being, it is
it is not a surprise that the spectrum is
shifted upward, because of the explicit hard-core specification of the
magnetic impurities.

So far, one has considered the average magnetic field limit $\alpha\to 0$.
Clearly,  the analysis for intermediate values of $\alpha$ becomes
quite involved. However, in the
$\alpha=\pm 1/2$ case, one can explicitly test the effect of
the random distribution of vortices.
Most probably, for intermediate value of $\alpha$, the main features of
this
analysis should remain valid.
(\ref{PART'}) now reads
\be\label{pi} Z=Z_o<e^{-2<B>{\cal{S}}}>_{\{C\}}\ee
where ${\cal{S}}=\sum_{n\quad  odd}{S_n/ \pi}$
is a well-behaved random variable which, because
of the probability distribution
law for
the $S_n$'s,  scales like
$t$. Thus  the  $y=
\pi{\cal{S}}/t$ probability distribution
$P(y=\pi{\cal{S}}/t)$ can be obtained by
simulations on a lattice, where a number of steps ranging from 2000 to
32000 has been used.
The average level density is obtained by inverse
Laplace transform of (\ref{pi})
\be \label{rho}<\rho(E)>=\rho_o(E)\int_0^{{E\over 2\rho}}P(y=
\pi{\cal{S}}/t) )dy\ee
In Figure 1, $<\rho(E)>$ displays
 a Lifschitz tail at the bottom of the spectrum (around
$E\simeq 2\rho<Y>=\pi\rho/4$), where a behavior
$<\rho(E)>\simeq \exp(-\rho/E)$ is expected. The  energy level depletion
at the bottom of the spectrum is  coherent with the positive shift (\ref{sh})
in the
Landau spectrum of the average magnetic field.
It is also reminiscent of
the singular $A-B$ density of states depletion $\rho(E)-\rho_o(E)={
\alpha(\alpha-1)\over 2}\delta(E)$ \cite{deplet} ($0<\alpha<1$ is understood).
Interesting enough, (\ref{rho}) shows that $<\rho(E)>$ is a
function of $E/\rho$ only. In fact, this feature happens to be true for any
value
of $\alpha$, and in the presence of an external  magnetic field $B$ as well.
It is easy to convince oneself that the inverse Laplace
transform $<\rho(E)>$ of the partition function $Z_B$ given in (\ref{SA})
is necessary a function of $E/\rho, B/\rho$ and $\alpha$.
This is precisely what is found next, in a quantum
mechanical approach.

{\bf Quantum Mechanics :}
One wishes to get some information on the
broadening of the Landau levels, and also on
the precise origin of the shift $<\omega_c>$ in the Landau
spectrum of the average magnetic field. The Hamiltonian of an electron in
a constant magnetic field $\vec B
=B\vec k$ ($|\alpha|<1$, $eB>0$ has been assumed without loss of generality),
superposed
onto $N$ vortices ($\vec k$ is the unit vector perpendicular to the plane)
reads
\be H={1\over 2m}\left(\vec p - \sum_{i=1}^N\alpha{\vec k\times(\vec r-\vec
r_i)
\over (\vec r -\vec r_i)^2}-{eB\over2} \vec k\times\vec r\right)^2\ee
In the presence of $B$, the sign of $\alpha$ becomes a physical
observable, i.e. the interpolations
$\alpha : -1/2\to 0$ and $\alpha : 0\to 1/2$
are not symmetric with respect to $\alpha=0$, or, in other terms, the
limits $\alpha\to 0^+$  or $\alpha\to 0^-$ differ.
In order to illustrate this statement, consider the degenerate
ground state (energy $\omega_c=eB/2m$) \cite{these}
\be \label{GS}\psi_o(\vec r)=\prod_i|z-z_i|^{m_i-\alpha}e^{im_i\arg(z-z_i)}
e^{-{1\over 2}m\omega_c z\bar z}\quad m_i\ge\alpha\ee
The eigenstates (\ref{GS}) vanish
when the electron coincide with a magnetic impurity. This amounts
to a particular choice of self-adjoint extension in the $m_i=0$,
$(\alpha<0)$ or
$m_i=1$, $(\alpha>0)$ angular
momentum sectors, where one could as well have normalisable but
diverging eigenstates at coinciding points.  This restriction to vanishing
functions
at the location of the vortices means unpenetrable
vortices, a hard-core boundary prescription that has already been discussed
in the brownian motion approach, where it was  related to the constant
positive shift
in the Landau
spectrum.
In a Hamiltonian formulation, it is given a precise meaning
by adding to $H$ contact repulsive
interactions, $H\to H'=H+\sum_i {\pi|\alpha|\over m}\delta^2(\vec r-\vec r_i)$,
which amount to couple the
spin 1/2 degree of freedom \cite{stefan} of the test particle,
endowed with a magnetic moment
$\mu=-{e\over 2m}\alpha/|\alpha|$ (thus an electron with gyromagnetic
factor $g=2$),  to the infinite magnetic field
inside the flux-tubes.

An important consequence is that the ground state basis (\ref{GS}) is complete
when $\alpha<0$, since
one obtains the complete LLL basis of the $B$ field in the limit $\alpha \to
0^-$ $(m_i\ge 0$).
 On the contrary, when
$\alpha \to 0^+$ $(m_i>0)$, $N$ excited states, which are not
analytically known, merge in the ground state to yield a complete
LLL basis \cite{these}. Clearly, when $\alpha\to 1$, these excited states
have to become usual excited Landau levels, simply because the system is
periodic. This pattern for the excited states is crucial for the
analysis that follows.

To proceed further, one takes into account the
behavior of the ground state near the
magnetic
impurities, quite analogously to what is done
in the
$A-B$ and anyon perturbative analysis \cite{acad}.
In the present situation, however, one is rather interested by
averaging over the disorder, and by properly reproducing the
average magnetic field contribution to the Landau spectrum of the test
particle. One performs the nonunitary
transformation
\be \label{new}{\psi}(\vec r)
=
e^{-{1\over2}m<\omega_c>r^2}\prod_i\vert\vec r-\vec r_i\vert^{\vert\alpha\vert}
\tilde{\psi}(\vec r)\ee
to obtain an Hamiltonian $\tilde{H}$ acting on $\tilde{\psi}(\vec r)$
where the impurity potential reads
\be \label{200}{V}(\alpha<0)
=\sum_{i=1}^N{2\alpha\over m}{\partial_{\bar z}
\over z-z_i}
+\sum_{i=1}^N(\omega_c+<\omega_c>)\alpha{z\over z-z_i}\ee
\be\label{100} {V}(\alpha>0)
=-\sum_{i=1}^N{2\alpha\over m}{\partial_z
\over\bar z-\bar z_i}
+\sum_{i=1}^N(\omega_c+<\omega_c>)\alpha{\bar z\over \bar z-\bar z_i}\ee
The average magnetic field exponential factor in (\ref{new}) has to be
understood in
 the infinite density
limit as compensating for
$\prod_i{\vert\vec r-\vec r_i\vert^{\vert\alpha\vert}} =
e^{\sum_{i=1}^N|\alpha|\ln\vert\vec r-\vec r_i\vert}\to
e^{\rho|\alpha|\int d^2\vec r'\ln\vert\vec r-\vec r'\vert}
=e^{{1\over2}m<\omega_c>r^2}$.
This is precisely the canonical transformation which factorizes
the standard pre-exponential factor in the Landau Hilbert space of the
average $<B>$ field. Short range singular interactions
$\alpha^2/|z-z_i|^2$ present in $H'$ have
been traded off for regular interactions $|\alpha|\partial_{\bar
z}/(z-z_i)$ in $\tilde{H}$.
Moreover, interactions involving
two magnetic
impurities have disappeared from $\tilde{H}$, which
will  greatly simplify the average on the disorder.

The Hamiltonians $H'$ and $\tilde H$ are equivalent, and can be
indifferently used for computing the partition
function or the density of states.
More rigorously, consider instead of (\ref{new})
\be \label{new'}\psi(\vec r)
=\prod_i{\vert\vec r-\vec r_i\vert^{\vert\alpha\vert}
\over <\vert\vec r-\vec r_i\vert^{\vert\alpha\vert}>}
\tilde{\psi}(\vec r),\ee
where the average $<>$ is done in a finite volume $V={N/\rho}$.
If the redefinition
(\ref{new'}) affects the short distance behavior of the Hilbert space, it
does not modify the long distance behavior of the wave functions, and
thus their normalisation.
It
leads to
a redefined Hamiltonian $\tilde{H}$ which is appropriate for estimating
$<\rho(E)>$ in a functional approach \cite{bre}. In the thermodynamic limit
$N\to\infty, V\to\infty$, the average density obtained in this way can
be shown \cite{preparation} to coincide
with the one derived from (\ref{new}, \ref{100}, \ref{200}).

If one  now extracts the mean-value  of ${V}(\alpha)$, one obtains
\be \label{1000}\tilde{H}= <\omega_c>+H_{B+<B>}+
{V}(\alpha)-<{V}(\alpha)>\ee
where $H_{B+<B>}
=-{2\over m}\partial_z\partial_{\bar z}+{m\over 2}{\omega_t}^2
z\bar z
-{e(B+<B>)\over 2m}(z\partial_z-\bar z\partial_{\bar z})$
is the Landau
Hamiltonian for the $B+<B>$ field. One has added for convenience to
$\tilde{H}$
a long distance harmonic
regulator $m\omega^2r^2/2$, which partially lifts the
degeneracy of the spectrum -one has
$\omega_t^2=({e(B+<B>)\over 2m})^2+\omega^2$,
the thermodynamic limit is obtained by
letting
$\omega\to 0$.  The global
shift $<\omega_c>$ appears explicitly in (\ref{1000}), in the average
field limit
$\alpha\to 0, \rho\to \infty$, where $V(\alpha)-<V(\alpha)>\to 0$.

If $\alpha<0$, $\tilde{H}$ should be trivially diagonal when
restricted to the
LLL of the $B+<B>$ field, since the redefined ground state basis
(\ref{new}) is  identical (the excited  states play no
role at all) to
the LLL basis $<u|z>=u(z)\exp(-m\omega_t
z{\bar{z}}/2)$, with $u(z)$ holomorphic\footnote{One assumes
  $\omega_c$ larger than $<\omega_c>$. If not, an antiholomorphic basis
has to be used. However, one can as well consider the opposite
situation with $e<B>$ positive, and $eB$ negative, i.e.  $<B>+B >0$,
which is in
fact described by (\ref{ap}).}. One respectively finds, when $\alpha<0$ and
$\alpha>0$
\be\label{am} <{v}\vert  {V}(\alpha)-<V(\alpha)>\vert u>=
<{v}|\sum_{i=1}^N(\omega_c-\omega_t)\alpha z({ 1\over z-z_i}-\pi\rho \bar
z)|u>\ee
\be\label{ap} <{v}\vert  {V}(\alpha)-<V(\alpha)>\vert u>=
<{v}|\sum_{i=1}^N(\omega_c-\omega_t)\alpha\bar z({1\over \bar z-\bar
z_i}-{\pi\rho} z)
+{2\pi\alpha\over m}(\sum_{i=1}^N\delta(z-z_i)-\rho)|u>\ee
In the
thermodynamic limit
$\omega\to 0$,
(\ref{am}) vanishes, so that
$\tilde{H}=\omega_c$ is indeed diagonal on the LLL basis.
More
interesting is the case
 $\alpha>0$, since then
\be\label{delta} \tilde{H}=<\omega_c>+H_{B+<B>}+{2\pi\alpha\over m}
(\sum_{i=1}^N \delta (z-z_i)-\rho)\ee
Thus, the magnetic impurity problem, when projected on the LLL of $B+<B>$,
is mapped on a  $\delta$ impurity problem.
These $\delta$ interactions
materialize the effect of the $N$ excited
states that leave the ground state when $\alpha\to 0^+$,
a fact that will be manifest in the average
density of states.
{}From this respect, strict periodicity is
lost since the analysis in the case  $\alpha>0$ is richer than in the
case $\alpha<0$.

This contact interactions  pattern happens to be
valid for vortex systems in general, when the excited states which join the
ground state are properly taken into account. It holds in particular for
the $N$-anyon Hamiltonian $\tilde{H}_N$ which, when projected on the LLL of an
external $B$
field,  becomes $\tilde{H}_N=\sum_{i=1}^N H_B(\vec {r}_i)+{4\pi\alpha\over
m}\sum_{i<j}\delta(\vec {r}_i-\vec {r}_j)-(\omega_t-\omega_c){N(N-1)\over 2}
\alpha$.

Projecting on the LLL of a Landau basis is meaningful if
one has a magnetic
field at disposal, which builds a well separated Landau spectrum for the
test particle. When the external magnetic field is
strong compared
to the average magnetic field, namely
$\rho_L>>\rho|\alpha|$,
one is certain that
the $V(\alpha)-<V(\alpha)>$ contribution
is small compared to the cyclotron gap.
On the other hand, in the regime where the
external magnetic field is comparable or smaller
than $<B>$, a physical  average
magnetic field is needed, namely $\rho\to\infty,\alpha\to 0$,
with $\rho\alpha$
finite. In a semi-classical point of view, the number
$f=\rho/(\rho_L+\rho\alpha)$
of magnetic impurities
enclosed by the classical electronic orbit in
the magnetic field $B+<B>$ has to be
big ($\rho_L=eB/2\pi$ is the Landau degeneracy). If this is the case,
${V}(\alpha)-<{V}(\alpha)>$ does not couple
different Landau levels of $B+<B>$.

What is the effect of disorder ?  One is
interested by the average density of states, a local quantity, which can
be directly computed in the thermodynamic limit $\omega=0$. If $\alpha<0$, the
impurity potential
vanishes, and the disorder has no effect on the  physical observable.
On the other hand, when $\alpha>0$,
one has to average on $\delta$ contact interactions, which
encode  non trivial effects due to the magnetic impurities. Such a $\delta$
impurity problem was studied by Brezin et al \cite{bre} -motivated by an
original study of  Wegner \cite{wegner}- using
supersymetric functional methods. They considered
a Poisson distribution (it is nothing but (\ref{prob}) in the
thermodynamic limit) of uniformly distributed $\delta$
impurities
in the LLL of a ${\cal{B}}$ field, with Hamiltonian $H=H_{{\cal{B}}}
+\lambda\sum_{i=1}^N\delta(\vec r-\vec r_i)$.
If one now sets $\lambda=2\pi\alpha/m$,
${\cal{B}}=B+<B>$,   $\nu={f\over
\lambda\rho}(E-\omega_c)$, one finds \cite{bre} that the average
density of states of (\ref{delta}) reads
\be\label{fini}<\rho(E)>={1\over\pi\lambda}Im\quad \partial_{\nu}\ln\int_o
^{\infty}dt\exp(i\nu
t-\epsilon t-f\int_o^t{dt'\over t'}(1-e^{-it'}))\ee
(the limit $\epsilon\to 0$ is understood). It
is transparent from (\ref{fini}) that $<\rho(E)>$ is
 a function of $\alpha$, $B/\rho$ and $E/\rho$, as already emphasized in
the brownian motion analysis.

{\bf Discussion and Conclusion :}
If $\alpha<0$, one trivially gets that $<\rho(E)>$ is a $\delta$ peak
centered at
$E=\omega_c$ with degeneracy $\rho_L+\rho\alpha$ (remember that when
$\alpha<0$,
$\omega_c$ has been assumed to be larger than $<\omega_c>$). A direct
computation, based on (\ref{am}),
 of the partition function of the
test particle in the presence of a harmonic regulator,
leads to the same result, once the thermodynamic limit is taken
\cite{preparation}.

For $\alpha>0$,  interesting  effects due to disorder are
expected. In the regime where the external magnetic field is dominant,
where typically $f<1$, the results read off \cite{bre} give a
$\delta$ peak also centered at $E=\omega_c$, with the same (by periodicity)
degeneracy
$(\rho_L+\rho\alpha)(1-f)= \rho_L
+\rho(\alpha-1)$ and $\rho$ additional states broadening the
LLL as $(E-\omega_c)^{-f}$. These states exactly correspond in the
thermodynamic limit to
the $N=\rho V$ excited states leaving  the grounsdate.
$f=1$ is critical simply because
since the LLL is then entirely
depleted by the excited states (the degeneracy of the LLL grows as
$\rho\alpha$, but the depletion grows as $\rho$). One can also easily
understand the bump in the density of states observed at $\nu=1$, i.e. at
energy
$E=\omega_c+{2\pi\alpha\over
m}(\rho_L+\rho\alpha)=\omega_c(1+2\alpha+2\alpha^2{\rho/\rho_L})$,
as the excited states contribution to the density of states. Indeed,
when $\alpha\to 1$, where a usual Landau spectrum has to be recovered,
one finds $E=3\omega_c$, to a correction of order $\rho/
\rho_L$, a manifestation of the approximation made when projecting on
the LLL (note that when $\alpha\to 1$, one necessery  has
$\rho_L>>\rho\alpha\simeq \rho$, thus $\rho/\rho_L\simeq 0$). So, most of the
$\rho$
excited states join the first excited Landau level of the $B$ field when
$\alpha=1$ (the others excited states join the Landau spectrum at higher
levels).

In the average magnetic field regime, where $f$ has to be big,
the shift
in the spectrum can be directly understood from (\ref{fini}).
One has $\nu\to\infty$, therefore
 only small values of $t$ contribute to the oscillating term,
implying that the
integral $\int
dt'(1-e^{-it'})/t'\simeq it$. One gets in the density of states the
shift $\nu\to\nu-f$, that is to say the desired $E\to E- 2\times
{e<B>\over 2m}$.
In the particular case $B=0$, the magnetic field is entirely due to the
magnetic impurities. Then, the parameter $f=1/\alpha$ depends only on
the $A-B$ coupling constant.

In conclusion, an  open question concerns the  conductivity \cite{kunz}
properties of an electron in the presence of magnetic impurities.
In the Quantum Hall regime, information about localisation
properties of the eigenstates would be of great interest.
Moreover, the test particle has been shown to satisfy a kind of
exclusion principle with the
magnetic impurities, via contact repulsive interactions.
It would be more satisfactory if statistical effects could be also
encoded in the distribution of the impurities themselves.
At last, concerning the brownian motion approach, the random magnetic field
 problem (windings) and the $\delta$ impurity problem (excursions)
 are, a priori, very different. The fact that they coincide when one
considers
their projection on a LLL certainly deserves further
explanations.

Acknowledgements : We acknowledge stimulating discussions with A. Comtet
and C. Monthus. J.D. acknowledges discussions with W. Werner and
S.O. with A. Dasni\`eres de
Veigy.

Figure captions :

Figure 1 : The average level density of states $<\rho(E)>$
as a function of the
variable ${E/(2\rho)}$ ($\alpha=\pm 1/2$, no external field)
exhibits a Lifshitz tail at the bottom of the spectrum.

\end{document}